\documentclass[a4paper,twocolumn,fleqn]{article}

\usepackage{amsmath}             
\usepackage{txfonts}               
\usepackage{helvet}
\usepackage{bm}                  
\usepackage{cite}                  
\usepackage{graphicx} 
\usepackage{pfr}                   

\newcommand{\const}{{\rm const}}

 \newcommand{\eq}[1]{Eq.~(\ref{#1})}
 
 \newcommand{\fig}[1]{Fig.~\ref{#1}}

\begin{document}

\title{Adiabatic Wave-Particle Interaction Revisited}

\author{Robert~L.~DEWAR\sup{1,2} and Justin~C.-C.~YAP\sup{1}
}

\affiliation{
\sup{1}Plasma Research Laboratory and Dept. of Theoretical Physics, Research School of Physics \& Engineering, The Australian National University, Canberra, ACT 0200, Australia
\\
\sup{2}Visiting Professor: Department of Advanced Energy, Graduate School of Frontier Sciences, University of Tokyo, Kashiwa-City
Kashiwanoha 5-1-5 Chiba 277-8561, Japan}

\date{Submitted 2008-10-31; published 2009-01-22; typos in Eqs. (8) \& (9) corrected 2009-01-26}

\email{robert.dewar@anu.edu.au}

\begin{abstract}
In this paper we calculate and visualize the dynamics of an ensemble of electrons trapping in an electrostatic wave of slowly increasing amplitude, illustrating that, despite disordering of particles in angle during the trapping transition as they pass close to X-points, there is still an adiabatic invariant for the great majority of particles that allows the long-time distribution function to be predicted. Possible application of this approach to recent work on the nonlinear frequency shift of a driven wave is briefly discussed.
\end{abstract}


\maketitle  

\section{\label{sec:0}Introduction}
The Bernstein-Greene-Kruskal (BGK) \cite{Bernstein_Greene_Kruskal_1957} Vlasov construction shows that fully nonlinear collisionless electrostatic waves and shocks, time-independent in some frame of reference, can in principle exist, but makes no statement as to the physical accessibility of these structures. The physical question is, given an initially time-independent, spatially uniform distribution function, which evolves into a nonuniform state either due to instability or external forcing, what is the final long-time steady state after coarse graining/phase-mixing the distribution function? (Also, in which frame is it time-independent?)

Consider a charged particle moving in a nonlinear wave propagating in the $z$-direction with electrostatic potential in the laboratory frame given by
\begin{equation}
	\phi(z,t) = u(\theta|\epsilon t) \;,
	\label{eq:generalwave}
\end{equation}
where $u(\theta)$ is $2\pi$-periodic, $u(\theta +2\pi) = u(\theta)$. In many cases $u$ can also be assumed to have half-period antisymmetry, $u(\theta +\pi) = -u(\theta)$ (i.e. it has only odd harmonics) and we shall assume this. The $\epsilon t$ dependence expresses the possibility that the amplitude and waveform can evolve slowly with time, the adiabatic limit being defined by $\epsilon \rightarrow 0$.

The phase angle $\theta$ is defined by
\begin{equation}
	\theta \equiv kz - \!\!\int^t_{0}\omega(\epsilon t')dt' \;,
	\label{eq:thetadef}
\end{equation}
where the wave vector $k{\bf e}_z$ is constant, but we have allowed for $\omega$ to evolve with time, e.g. due to a nonlinear frequency shift. The wave phase $\theta$ at the particle position forms a convenient generalized coordinate for describing the particle dynamics as it is nondimensional. Also, $\dot{\theta}/k = v_z - v_{\rm ph}$ is the particle velocity in the \emph{wave frame}, i.e. after a Galilean transformation to a frame moving at the phase velocity  $v_{\rm ph} \equiv \omega/k$. (Note that the wave frame is noninertial if $\omega$ is time-dependent.)
The equation of motion for a particle of mass $m$, charge $q$ is
\begin{equation}
	\ddot{\theta} = -\frac{qk^2}{m}\frac{\partial\phi}{\partial\theta} - \frac{d\omega}{dt} \;.
	\label{eq:eqmotion}
\end{equation}

The particle dynamics during the evolution between the initial and final states cannot be solved analytically because the energy is not a constant of the motion. However, Dewar \cite{Dewar_72b,Dewar_73a} showed that the approximation of adiabatic invariance, combined with phase mixing, provided a sufficiently accurate mapping between the initial and final distribution functions that reasonable estimates of nonlinear frequency shift and saturation amplitude of an unstable Langmuir wave could be made analytically.

Motivated by applications in laser fusion and beam physics, the utility of the adiabatic approximation has recently been confirmed numerically by Lindberg \emph{et al.} \cite{Lindberg_Charman_Wurtele_07} for a \emph{driven} Langmuir wave. In \cite{Lindberg_Charman_Wurtele_07} the long-time wave
response was calculated both by particle simulation and
semi-analytically using the adiabatic Vlasov approximation \cite{Dewar_72b,Benisti_Gremillet_07}
or the electron dynamics in a wave field of slowly varying
amplitude and frequency.  The authors allowed for a nonsinusoidal
waveform due to excitation of harmonics at large amplitudes, and for
the generation of a dc electric field to satisfy the assumed external
circuit conditions of zero initial and final spatial-mean current
(i.e. the dc component of the current).  Both methods of calculation
agreed well, confirming the utility of the adiabatic approximation.

In the small-amplitude limit, where the waveform is sinusoidal,
Lindberg \emph{et al.} \cite{Lindberg_Charman_Wurtele_07} also calculated a nonlinear frequency shift
analytically in terms of elliptic integrals, finding an expression proportional to the
square root of the final wave amplitude, with a coefficient in
agreement with the earlier calculation by Dewar \cite{Dewar_72b}.  However, Fig.~4 of \cite{Lindberg_Charman_Wurtele_07} shows quite
poor agreement between this small-amplitude formula and their
finite-amplitude simulations and semi-analytic calculations, except at
very low amplitudes.

In Sec.~\ref{sec:adtrap} we specialize to a sinusoidal wave of fixed frequency and describe a numerical experiment in which a set of electrons of the same initial wave-frame energy, but with different phases, is evolved as the amplitude is increased. A special choice of growth function allows the extreme adiabatic limit to be probed and it is shown that adiabatic theory is statistically very accurate in this limit, even though particles originally launched within one wavelength of each other end up being trapped in several wave troughs. This is investigated in more detail in Sec.~\ref{sec:adtrapdetail} where the details of the trapping process are visualized, illustrating that close encounters with hyperbolic X-points can break up the ordering of the particles.

Section~\ref{sec:ensembletrap} shows the result of evolving an ensemble of particles, some of which are trapped while others remain free, illustrating why the coarse-grained distribution function of trapped particles is half its initial pre-trapping value. This is used in Sec.~\ref{sec:ftrans} to explain the basis of our earlier formalism \cite{Dewar_72b,Dewar_73a}.

In the concluding remarks in Sec.~\ref{sec:Conclusion} we mention some other early theoretical work \cite{Dewar_Lindl_72,Dewar_72a} that may point the way to improving the agreement between the asymptotic  amplitude expansion for the nonlinear frequency shift and the simulation results of Lindberg \emph{et al.} through the inclusion of higher order terms.

\section{\label{sec:adtrap}{Adiabatic trapping}}

To understand adiabatic trapping more clearly, in this section we study a specific example, a set of electrons with charge $q = -e$ moving in a sinusoidal wave
\begin{equation}
	\phi(z,t) = \phi_1(t)\cos\theta\;.
	\label{eq:sinusoid}
\end{equation}
We here assume $\omega$ constant, so $\theta = kz - \omega t$.

\begin{figure}[htbp]
	\centering
	 \includegraphics[width=6cm]{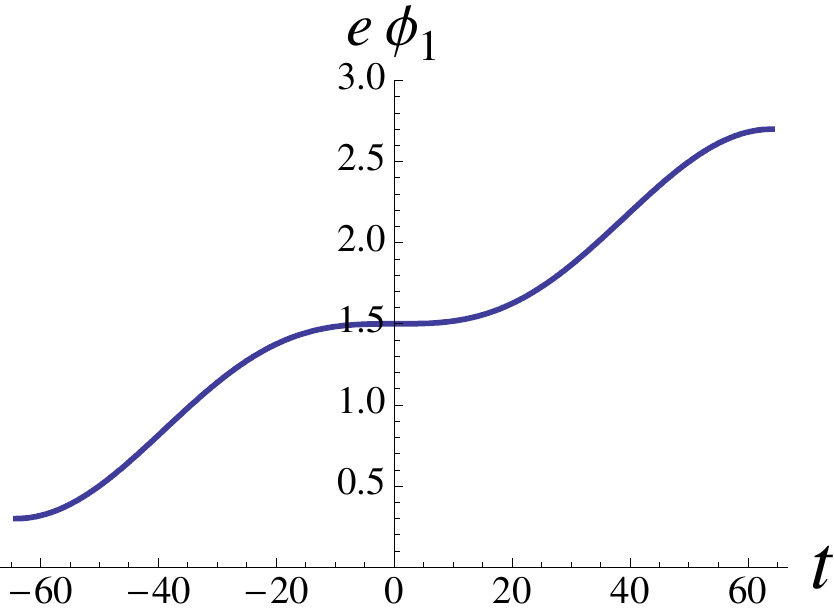}
	\caption{Adiabatically increasing amplitude of the electron potential energy, $e\phi$, in the wave \emph{vs.} time $t$, between $t = -t_{\rm f}$ and $t = t_{\rm f}$ as described in the text.}
	\label{fig:adiabaticramp}
\end{figure}

In the autonomous case, $e\phi_1 = \const$, the wave-frame energy
\begin{equation}
	W \equiv \frac{m}{2k^2}\dot{\theta}^2 + q\phi
	\label{eq:Wdef}
\end{equation}
is a constant of the motion, which allows solution of this one-degree-of-freedom system by quadratures. For the sinusoidal wave there is a direct analogy with the physical pendulum, where $\theta$ is the angle relative to the vertical. The motion of an autonomous nonlinear pendulum is known to be solvable in Jacobian elliptic functions \cite{Abramowitz_Stegun_72}. For a pendulum of length $l$, the analogue of $ek^2\phi_1/m$ is $g/l$. Thus, for the pendulum the analogue of a changing wave amplitude would be a changing gravitational field $g(t)$. Trapping in a wave is analogous to the transition from rotation to libration for a pendulum.

In this section we investigate the dynamics of a set of initially untrapped electrons moving in a wave potential of the form specified in \eq{eq:sinusoid}, with the amplitude function
\begin{equation}
	\label{eq:rampfn}
	\phi_1 = \phi^0_1 + \frac{3\Delta\phi}{4}
	\left[ \sin \left(\frac{\pi}{2} \frac{t}{t_{\rm f}}\right) - \frac{1}{3} \sin\left(\frac{3\pi}{2} \frac{t}{t_{\rm f}}\right)\right]\,,
\end{equation}
and with $\phi^0_1 = 1.5$ and $\Delta\phi = 1.2$ in units such that $e = k = m =1$. This function, depicted in \fig{fig:adiabaticramp}, is chosen so that its time rate of increase is zero at the beginning.  $t = -t_{\rm f}$, middle, $t = 0$, and end, $t = t_{\rm f}$, of the calculation period. This choice makes the adiabatic description very accurate while keeping the total time of the calculation reasonably short: defining $t_{\rm f} = \pi/\epsilon\omega^0_{\rm b}$, where $\omega^0_{\rm b}\equiv k(e\phi^0_1/m)^{1/2}$ is the bounce frequency at the bottom of a wave potential trough at time $t = 0$, the results in this paper were obtained with $\epsilon = 0.04$, giving $t_{\rm f} = 64.13$.
\begin{figure}[htbp]
	\centering
	\includegraphics[width=6cm]{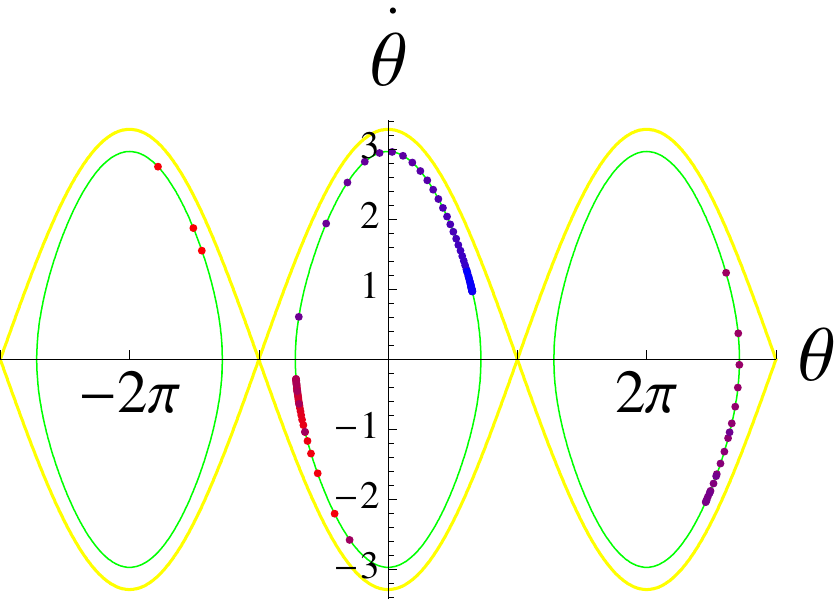}
	\caption{Phase-space positions of the set of 101 particles described in the text at the final time $t = t_{\rm f}$.}
	\label{fig:phase-trapping}
\end{figure}

Figure~\ref{fig:phase-trapping} shows the final positions of a set of 101 electrons initialized at $t = -t_{\rm f}$ with the same initial total energy, $W = 1.22515$, but at different initial points $(\theta,\dot{\theta})$ on the upper $W$-contour, i.e. with positive velocity $\dot{\theta}$. The particles were initialized in an interval of $\theta$ of width $2\pi$ whose endpoints were chosen to the left of the origin, such that all the particles trapped in wave troughs near the origin. Each electron was assigned a unique color, starting from red at the left initial endpoint to blue on the right initial endpoint.

Rather than using action-angle variables \cite{Lindberg_Charman_Wurtele_07} we adopt a more Lagrangian approach and study the particle orbits in $\theta,\dot{\theta}$ phase space. It is a standard result of adiabatic invariant theory that the phase-space area $\int\!\!\int d\theta d\dot{\theta}$ under the contour $W = C(\epsilon t)$, where $C$ denotes a constant of the motion in the autonomous case, is an adiabatic invariant for both trapped and passing (free) particles. However, Best \cite{Best_68}\footnote{See also Elskens and Escande \cite{Elskens_Escande_03} and references therein.} discovered that this is true for most particles \emph{even through the trapping or detrapping transition}, provided the $\theta$ integral is taken over one $2\pi$ period between the upper $W$-contour and the $\theta$-axis while the particle is free, or over \emph{half} a $W$-contour when the particle is trapped.

Following Dewar \cite{Dewar_72b,Dewar_73a} we divide the phase-space area by $2\pi$, so that the adiabatic invariant becomes the $\theta$-\emph{average} of the positive branch of the solution for the angular velocity of the equation $W = \const$,
\begin{equation}
	\overline{\dot{\theta}}(W) \equiv
	\left(\frac{2}{m}\right)^{1/2}\!\!\frac{k}{2\pi}\int_{-\pi}^{\pi}\!\!\!d\theta\,
										H(W-q\phi)\,(W-q\phi)^{1/2}
	\label{eq:AdAv}
\end{equation}
where $H(\cdot)$ is the Heaviside step function.

Equation~(\ref {eq:AdAv}) is valid as an adiabatic invariant for an arbitrary waveform as in \eq {eq:generalwave} provided the nonlinear frequency shift can be ignored. In the special case of a sinusoidal wave, the integration can be done analytically in terms of complete elliptic integrals $E(m)$ and $K(m)$ \cite{Abramowitz_Stegun_72}. For passing particles we have
\begin{equation}
	\overline{\dot{\theta}}(W) = \frac{4\omega_{\rm b}}{\pi} w^{1/2} E(w^{-1}) \;,
	\label{eq:AdAvfree}
\end{equation}
while for trapped particles,
\begin{equation}
	\overline{\dot{\theta}}(W) = \frac{4\omega_{\rm b}}{\pi} [E(w) - (1-w)K(w)] \;,
	\label{eq:AdAvtrapped}
\end{equation}
where $w(t,W) \equiv (W + e\phi_1)/(2e\phi_1)$ is $>1$ for passing particles and $< 1$ for trapped particles. Here $\omega_{\rm b}(t) \equiv  k(e\phi_1/m)^{1/2}$ is the bounce frequency at the bottom of a wave trough at time $t$.

In \fig{fig:phase-trapping} the final separatrix, i.e. the contour $W =  e\phi_1(t_{\rm f})$ passing through the X-points at odd multiples of $\pi$, is shown in yellow. Trapping comes about due to the expansion of the separatrix as $e\phi_1$ increases---when the area enclosed by the separatrix exceeds the area under one period of the upper initial $W$-contour the particles must trap to conserve the adiabatic invariant. This was used to choose the initial value of $W$ so that the adiabatic theory prediction for the trapping time was $t = 0$.

\begin{figure}
	\centering
	\includegraphics[width=6cm]{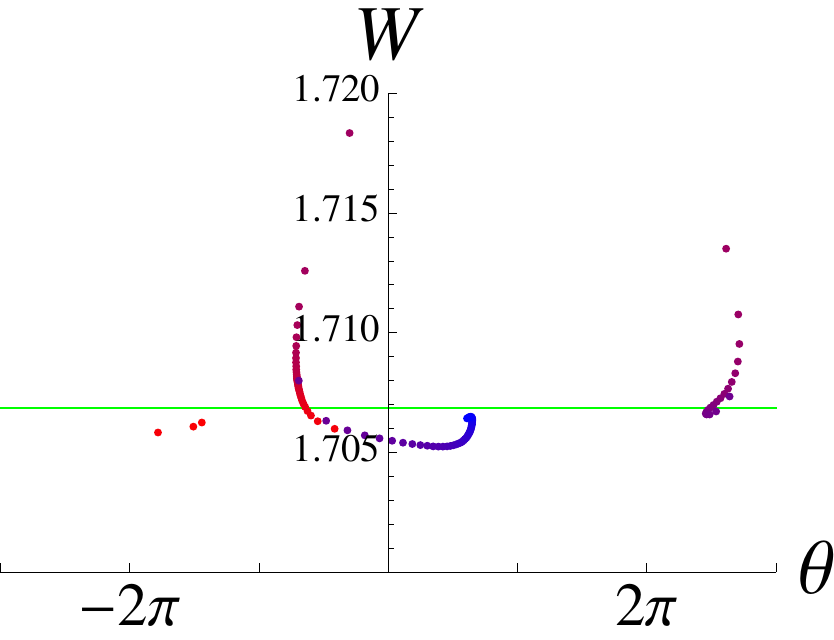}
	\caption{Particles from the initially well-ordered set are, at $t = +t_{\rm f}$, distributed over three different wave troughs but have total energies very close to the adiabatic prediction.}
	\label{fig:posttrapzoom}
\end{figure}

\begin{figure}[htbp]
	\centering
	 \includegraphics[width=6cm]{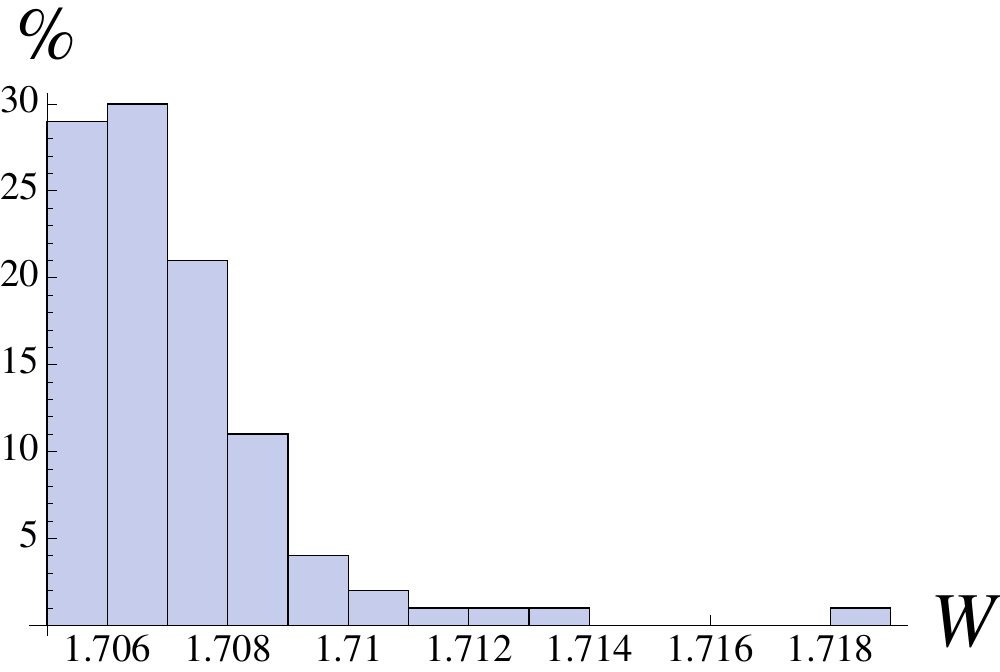}
	\caption{Histogram showing the statistical distribution of final energies of trapped particles. }
	\label{W-trapping}
\end{figure}

The adiabatic prediction for the \emph{final} energy is shown in green, and it is seen that, at the resolution of the plot, all the particles do lie on the contour predicted by Best's theory, but the higher resolution $W$-$\theta$ representation in \fig{fig:posttrapzoom} shows departures from adiabaticity. However, even the least adiabatic points are quite close to adiabatic and most points cluster close to the adiabatic prediction. 

In \fig{W-trapping} we quantify this observation of clustering, verifying that the great majority of particles do lie close to the adiabatic prediction. The mean energy at $t = +t_{\rm f}$ is $\overline{W} = 1.7072$ with a standard deviation $0.002$, which is $0.1\%$ of the mean. However, the standard deviation greatly overestimates the discrepancy between $\overline{W}$ and the adiabatic prediction, $W_{\rm ad} = 1.70687$, which is within $0.02\%$ of the mean.

\begin{figure}[htbp]
	\centering
	 \includegraphics[width=6cm]{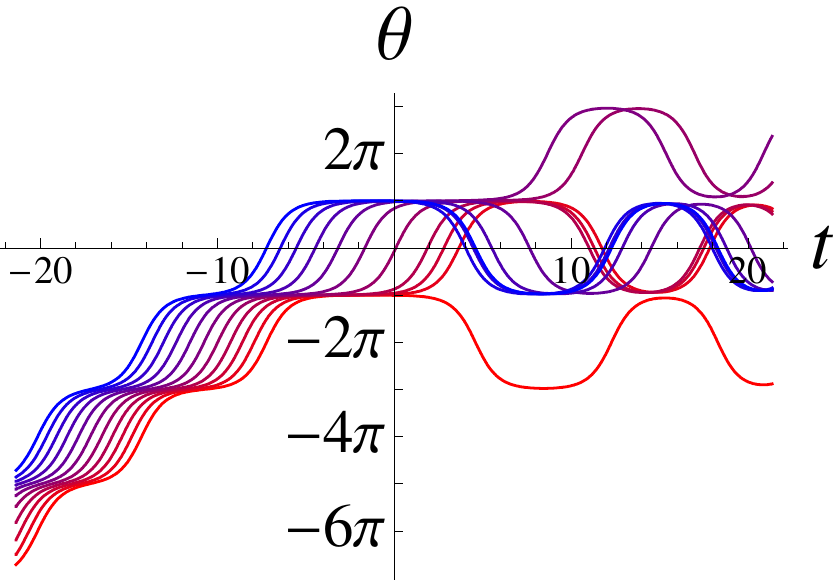}
	\caption{Onset of trapping in angle space for a subset of 11 particles starting on an energy contour outside the separatrix with a spread of $2\pi$ in angle, but which trap in three different wave troughs.}
	\label{fig:ang-trapping}
\end{figure}

\section{\label{sec:adtrapdetail}{Details of trapping}}
We see in Figs.~\ref {fig:phase-trapping} and \ref {fig:posttrapzoom} that the initial ordering in $\theta$ has been severely disrupted by the final time, with the particles, initially within the same $2\pi$ interval in $\theta$, now spread over three different wave troughs. To visualize the mechanism for this disordering we now examine the trapping transitions more carefully.

\begin{figure}[htbp]
	 \includegraphics[width=6cm]{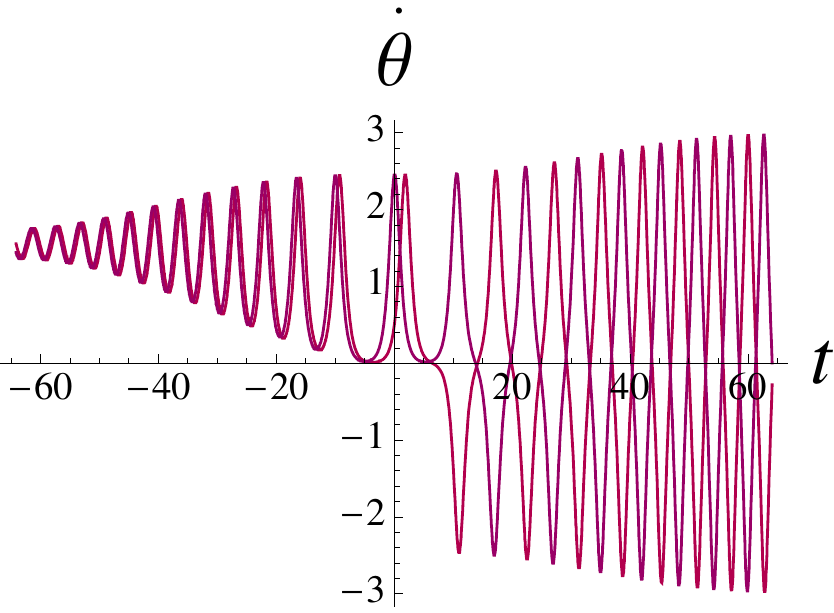}
	\centering
	\caption{Angular velocity for two adjacent orbits from the set used in \fig{fig:ang-trapping}, which trap in different wave troughs. The mean angular velocity evolves slowly until rapidly dropping to zero at the onset of trapping.}
	\label{fig:angfreq-trapping}
\end{figure}

As is seen in Figs.~\ref{fig:ang-trapping} and \ref{fig:angfreq-trapping}, the particles do indeed pass through the separatrix and become trapped in the vicinity of $t = 0$, when the rate of increase of the amplitude is very small. Because of this, the \emph{effective} $\epsilon$ is much smaller during the trapping period (when adiabatic theory is least accurate) than its actual value of $0.04$. Comparing the slope calculated at a typical trapping time (taken to be $\sim 5$ on the basis of the results shown in \fig{fig:ang-trapping}), of a simpler ramp-up function, $\phi^0_1 + \Delta\phi\sin \pi t/2 t_{\rm f}$, we estimate the effective adiabatic expansion parameter to be $\epsilon_{\rm eff} \sim 2 \times 10^{-3}$.

\begin{figure}
	\centering
	\includegraphics[width=6cm]{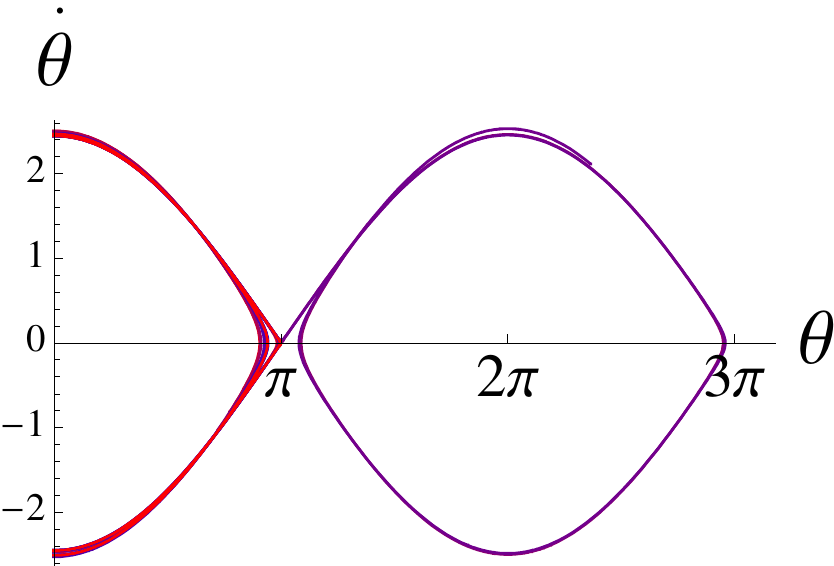}
	\caption{Phase-space plots of the trapping of the set used in \fig{fig:ang-trapping} showing the two trapping troughs to the right of the $\theta$ origin.}
	\label{fig:trappingphasespace}
\end{figure}

\begin{figure}
	\centering
	\includegraphics[width=6cm]{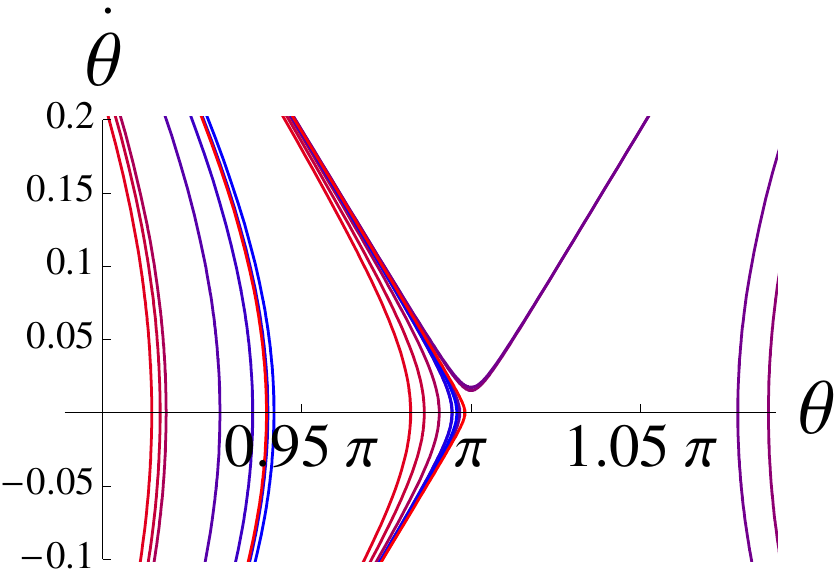}
	\caption{Zoomed plot of the set used in \fig{fig:trappingphasespace}.}
	\label{fig:trapphspacezoom}
\end{figure}

It is also instructive to examine the trapping process in phase space. We see in Figs.~\ref{fig:trappingphasespace} and \ref{fig:trapphspacezoom} that all particles come very close to the hyperbolically unstable X-points before they finally trap, thus amplifying the small differences in the individual histories of each orbit and accounting qualitatively for the disordering in $\theta$ seen at the final time in Figs.~\ref {fig:phase-trapping} and \ref {fig:posttrapzoom}. We now see that this disordering occurs during trapping and is presumably not primarily due to the accumulation of phase differences during the subsequent spiraling in toward the final adiabatic energy, though the small deviations from this energy seen in \fig{fig:trapphspacezoom} will contribute to disordering from small variations in the nonlinear bounce frequencies.

\begin{figure}
	\centering
	\includegraphics[width=6cm]{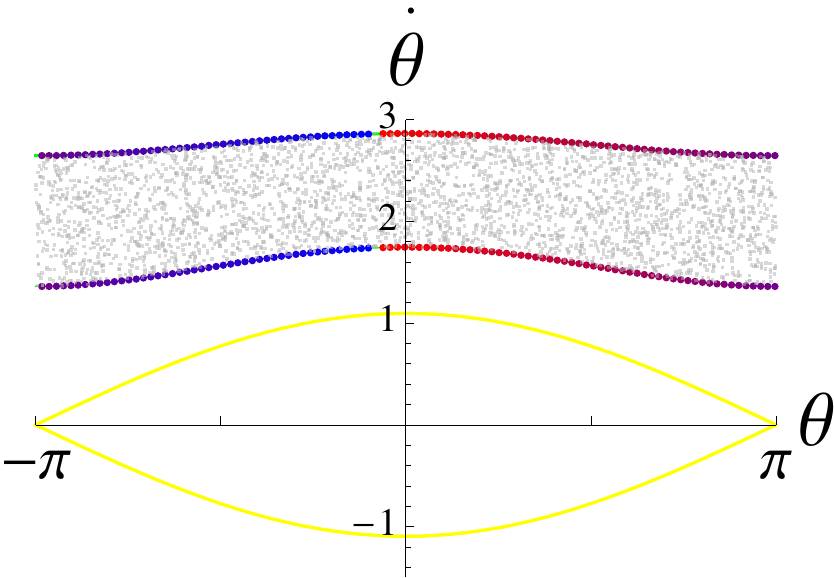}
	\caption{Initial state of ensemble, at $t = -t_{\rm f}$.}
	\label{fig:pretrap}
\end{figure}

\section{\label{sec:ensembletrap}{Trapping of an ensemble}}

\begin{figure}
	\centering
	\includegraphics[width=6cm]{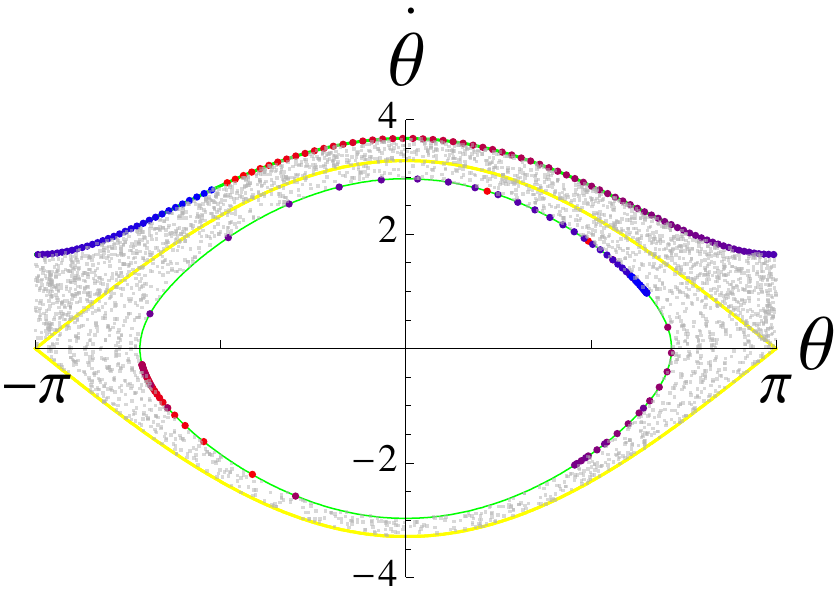}
	\caption{Final state of ensemble, at $t = +t_{\rm f}$.}
	\label{fig:posttrap}
\end{figure}

In \fig{fig:pretrap} we show an ensemble of initial points (in gray) randomly distributed with equal phase-space probability density, $f(\theta,\dot{\theta},-t_{\rm f}) = \const$, between two $W$-contours. The lower contour is the same one used in Sec.~\ref{sec:adtrap} above, with the same color-coded set of 101 initial points shown, but projected modulo $2\pi$ onto the interval $-\pi,\pi$. That is, we are now using the cylindrical topology of the pendulum phase space (or, alternatively, duplicating the initial set of points across all periods of the wave). The upper contour is chosen so that electrons on it never trap, and 101 color-coded initial points on this contour are also shown.

Figure~\ref{fig:posttrap} shows the final state after the points shown in \fig{fig:pretrap} are evolved dynamically from $t = -t_{\rm f}$ to $t = t_{\rm f}$. The trapped points on the inner adiabatic contour (green) are the same ones shown in \fig{fig:phase-trapping} but now projected into one period as explained above. The untrapped points outside the separatrix on the upper contour have evolved adiabatically to a very high accuracy as they never come near the expanding separatrix (yellow).

As the area of phase-space elements is preserved by the dynamics, the probability density in the untrapped region, between the upper separatrix and the upper contour, is the same as used initially, $f_{\rm pass}(\theta,\dot{\theta},t_{\rm f}) = f(\theta,\dot{\theta},-t_{\rm f})$. However the trapped points, between the inner contour and the \emph{entire} separatrix, are now distributed quasi-randomly over \emph{twice} the area they occupied initially (because the adiabatically conserved area is that of only the inner half-contour \emph{above} the $\theta$-axis). Thus the coarse-grained probability density of the trapped particles is \emph{half} the initial value:
\begin{equation}
	f_{\rm trap}(\theta,\dot{\theta},t_{\rm f}) = \frac{1}{2} f(\theta,\dot{\theta},-t_{\rm f}) \;.
	\label{eq:trappedf}
\end{equation}

\section{\label{sec:ftrans}Distribution function after adiabatic wave excitation}
Here we relate the above result to the formalism used in our earlier calculations \cite{Dewar_72b,Dewar_73a}. We work in $z,v_z$ phase space in the lab frame, so the relation to the wave frame must be made explicit. Thus \eq{eq:Wdef} is now written
\begin{equation}
	W = \frac{1}{2}m(v_z - v_{\rm ph})^2 + q\phi(z-v_{\rm ph}t,\varepsilon t) \;,
	\label{eq:Wdeflab}
\end{equation}
where $v_z \equiv \dot{z}$ and phase velocity $v_{\rm ph}$ may include a nonlinear frequency shift. Solving \eq{eq:Wdeflab} for $v_z$ we have two solutions, corresponding to particles going to the right or left in the wave frame, $v_z^{\pm} = v_{\rm ph} \pm u$, where
\begin{equation}
	u \equiv (2/m)^{1/2}[W -  q\phi(z-v_{\rm ph}t,\varepsilon t)]^{1/2} \;.
	\label{eq;udef}
\end{equation}

We assume there to be no wave at $t = -\infty$, so then $u$ reduces to $u_0(W) \equiv (2/m)^{1/2}W^{1/2}$ and the initial distribution function $f_0(v_z)$ can be written as a two-branched function of $W$, which we denote by
\begin{equation}
	F^{\pm}_{-\infty}(W)  \equiv  f_0(v_{\rm ph} \pm u_0(W)) \;.
	\label{eq:initialF}
\end{equation}
The long-time, coarse-grained (or \emph{phase averaged} \cite{ONeil_1965}) distribution function $f_{\infty}(z,v_z,t)$ will be independent of time in the wave frame, and so must also be a function of the constant of the motion $W$, which we denote by $F_{\infty}(W)$:
\begin{equation}
	F^{\pm}_{+\infty}(W) \equiv f(z,\pm|v_z|,t=+\infty)
	\label{eq:Finfinitydef}
\end{equation}
The goal of adiabatic theory is to find the transformation between $F_{-\infty}(W)$ (or $f_0$) and $F_{+\infty}(W)$.

Restricting to the case where the nonlinear frequency shift and dc field can be ignored, we use the adiabatic invariant $\overline{\dot{\theta}}(W)$ defined in \eq{eq:AdAv}, or, rather, $\overline{u}(W) \equiv \overline{\dot{\theta}}(W)/k$. We also see from the previous section that a transition from a passing orbit at $t = -\infty$ to a passing orbit at $t = +\infty$ preserves both the value of the distribution function and the direction of the wave-frame velocity (provided there is no intermediate trapping), while a transition from a passing orbit to a trapped orbit mixes initially left-going and right-going passing particles in the same band of $W$ and halves the value of their individual distribution functions,

Thus the transformation between $f_0$ and $F_{+\infty}$ is as summarized below.
\begin{itemize}
  \item Passing to passing:
\begin{equation}
	F^{\pm}_{+\infty}(W) = f_0(v_{\rm ph} \pm \overline{u}(W)) 
\end{equation}
  \item Passing to trapped:
\begin{eqnarray}
	F^{\pm}_{+\infty}(W) &=&  \frac{1}{2} \left[f_0(v_{\rm ph}-\overline{u}(W))
	\right.\nonumber\\ 
					    & &	\left.\mbox{} + f_0(v_{\rm ph}+\overline{u}(W)) \right] \;.
\end{eqnarray}
\end{itemize}

\section{Conclusion}\label{sec:Conclusion}
The numerical and graphical study presented here verifies the applicability of the adiabatic approximation to the calculation of the long-time coarse-grained distribution function of a plasma after the growth and saturation of a slowly growing instability or driven wave. Yet to be studied in similar detail is the effect of an external electric field, and of a frequency that changes with time, in order to verify the adiabatic theories presented in previous work \cite{Dewar_73a, Lindberg_Charman_Wurtele_07}.

Also awaiting further work is a careful matched asymptotic expansion using adiabatic theory to calculate the nonlinear frequency shift of a driven wave beyond the first $O(\phi_1^{1/2})$ term in order to compare with the numerical calculations of Lindberg {\emph et al.} \cite{Lindberg_Charman_Wurtele_07}. While the $O(\phi_1^{1/2})$ term due to trapped particles must dominate for very small amplitudes, higher powers of $\phi_1$ will become dominant at higher amplitudes, and it is clear this must be occurring for the parameters used in Fig.~4 of \cite{Lindberg_Charman_Wurtele_07}. 

The $O(\phi_1^2)$ frequency shift due to the nonlinear response of the bulk of the distribution function, which is not subject to particle trapping effects, was calculated earlier using a ``waterbag'' distribution function (effectively a fluid model) by Dewar and Lindl \cite{Dewar_Lindl_72} and using an averaged-Lagrangian oscillation-center kinetic method by Dewar \cite{Dewar_72a}. Winjum \emph{et al.} \cite{Winjum_Fahlen_Mori_07} have recently postulated that such a ``fluid'' nonlinear frequency shift can simply be superimposed on the $O(\phi_1^{1/2})$ frequency shift from trapped particles, but it seems \emph{a priori} quite possible that a careful asymptotic expansion will reveal terms at intermediate orders,  $O(\phi_1)$ and $O(\phi_1^{3/2})$.

\section*{Acknowledgments}
The first author (RLD) wishes to acknowledge the hospitality of Professor Z. Yoshida and colleagues in the Department of Advanced Energy, Graduate School of Frontier Sciences, The University of Tokyo, where this work was completed. The calculations were performed using \emph{Mathematica} \cite{Mathematica6}. 

\bibliographystyle{prsty}
\bibliography{RLDBibDeskPapers}
\end{document}